\documentclass[pra,aps,twocolumn,preprintnumbers,showpacs,superscriptaddress]{revtex4}

\usepackage{graphics,graphicx,epsfig,bm,amsthm,amsmath,amssymb,color,hyperref}
  
\begin{document}
\title{Solving Quantum Ground-State Problems with Nuclear Magnetic Resonance}

\date{\today}

\author{Zhaokai Li}
\thanks{These authors contributed equally to this work.}
\affiliation{Hefei National Laboratory for Physical Sciences at Microscale and Department of Modern Physics, University of Science and Technology of China, Hefei, Anhui 230036, People's Republic of China}

\author{Man-Hong Yung}
\thanks{These authors contributed equally to this work.}
\affiliation{Department of Chemistry and Chemical Biology, Harvard University, Cambridge MA, 02138, USA}

\author{Hongwei Chen}
\affiliation{Hefei National Laboratory for Physical Sciences at Microscale and Department of Modern Physics, University of Science and Technology of China, Hefei, Anhui 230036, People's Republic of China}

\author{Dawei Lu}
\affiliation{Hefei National Laboratory for Physical Sciences at Microscale and Department of Modern Physics, University of Science and Technology of China, Hefei, Anhui 230036, People's Republic of China}

\author{James~D.~Whitfield}
\affiliation{Department of Chemistry and Chemical Biology, Harvard University, Cambridge MA, 02138, USA}

\author{Xinhua Peng}
\affiliation{Hefei National Laboratory for Physical Sciences at Microscale and Department of Modern Physics, University of Science and Technology of China, Hefei, Anhui 230036, People's Republic of China}

\author{Al\'{a}n Aspuru-Guzik}
\email{aspuru@chemistry.harvard.edu}
\affiliation{Department of Chemistry and Chemical Biology, Harvard University, Cambridge MA, 02138, USA}

\author{Jiangfeng Du}
\email{djf@ustc.edu.cn}
\affiliation{Hefei National Laboratory for Physical Sciences at Microscale and Department of Modern Physics, University of Science and Technology of China, Hefei, Anhui 230036, People's Republic of China}


\begin{abstract}
Quantum ground-state problems are computationally hard problems; for general many-body Hamiltonians, there is no classical or quantum algorithm known to be able to solve them efficiently. Nevertheless, if a trial wavefunction approximating the ground state is available, as often happens for many problems in physics and chemistry, a quantum computer could employ this trial wavefunction to project the ground state by means of the phase estimation algorithm (PEA). We performed an experimental realization of this idea by implementing a variational-wavefunction approach to solve the ground-state problem of the Heisenberg spin model with an NMR quantum simulator. Our iterative phase estimation procedure yields a high accuracy for the eigenenergies (to the $10^{-5}$ decimal digit). The ground-state fidelity was distilled to be more than 80\%, and the singlet-to-triplet switching near the critical field is reliably captured. This result shows that quantum simulators can better leverage classical trial wave functions than classical computers.
\end{abstract}


\maketitle

\section{Introduction}
Quantum computers can solve many problems much more efficiently than a classical computer \cite{Ladd2010}. One general class of such problems is known as {\it quantum simulation} \cite{Feynman1982, Kassal2010}. In this class of algorithms, the quantum states of physical interest are represented by the quantum state of a register of controllable qubits (or qudits), which contains the quantum information of the simulated system. In particular, one of the most challenging problems in quantum simulation is the {\it ground-state preparation} problem \cite{gsp} of certain Hamiltonians, $H$, which can be either classical or quantum mechanical. Remarkably, every quantum circuit \cite{Kitaev2002}, and even thermal states \cite{Somma2007, Yung2011}, can be encoded into the ground state of certain Hamiltonians, and purely mathematical problems, such as factoring \cite{Peng2008}, can also be solved by a mapping to a ground-state problem.

On the other hand, the ground-state problem has profound implications in the theory of computational complexity \cite{Papadimitriou}. For example, finding the ground-state of a general classical Hamiltonian (e.g. the Ising model) is in the class of $\mathsf{NP}$ (nondeterministic polynomial time) computational problems, meaning that while finding the solution may be difficult, but verifying it is efficient when employing a classical computer. The Ising model with non-uniform couplings is an example of an $\mathsf{NP}$-problem (more precisely, $\mathsf{NP}$-complete)~\cite{Barahona1982}. The quantum generalization of $\mathsf{NP}$ is called $\mathsf{QMA}$ (Quantum Merlin Arthur) \cite{Kitaev2002}. In this class, the verification process requires a quantum computer, instead of a classical computer. An example of a problem in $\mathsf{QMA}$ is the determination of the ground-state energy of quantum Hamiltonians with two-body (or more) interaction terms \cite{Kempe2006}. So far, there is no known algorithm, classical or quantum, that can solve all problems efficiently in $\mathsf{NP}$ and $\mathsf{QMA}$.

Most of the problems in physics and chemistry, however, exhibit special structures and symmetries, that leads to methods for approximating the ground state with trial states $\left| {\psi _{T} } \right\rangle$ (or trial wave-function) possible. For example, in quantum chemistry \cite{Helgaker2000}, the Hartree-Fock mean field solution often captures the essential information of the ground state~$\left| e_0 \right\rangle$ for a wide range of molecular structures. However, the applicability of these trial states will break down whenever the fidelity,
\begin{equation}\label{def_fidelity}
F \equiv | {\langle {e_0 } | {\psi _{T} } \rangle } |^2 \quad,
\end{equation}
quantified by the square of the overlap between the trial state  $\left| {\psi _{T} } \right\rangle$ and the exact state $
\left| {e_0 } \right\rangle$, is vanishingly small. Specifically, if the fidelity of a certain trial state for a particular many-body problem is small, for example, about $F=0.01$, it might be considered as a ``poor" approximation to the exact ground state \cite{Kohn2005}, when used as an input state in classical computation. For quantum computing, however, the same trial state can be a ``good" input, as one only needs to repeat the ground-state projection algorithm, e.g., by Abrams and Lloyd \cite{Abrams1999} (see below), for about $O$(100) times, which is computationally efficient especially when the Hilbert space of the many-body Hamiltonian is usually exponentially large. This is the motivation behind our experimental work.

\begin{figure}[t]
\includegraphics[width=  0.9 \columnwidth]{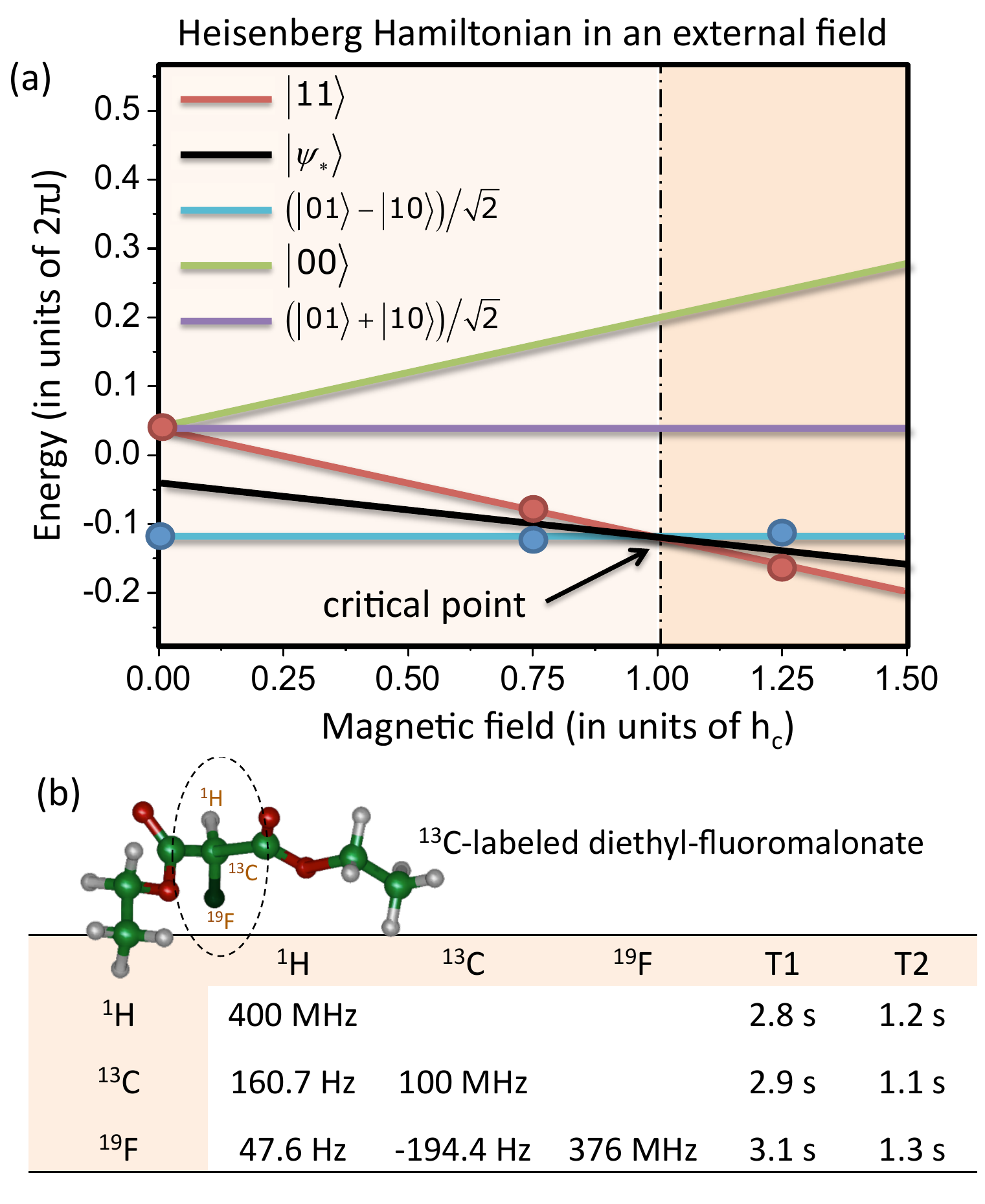}
\caption{(Color online) (a) The energy eigenvalues versus external magnetic field of the Heisenberg Hamiltonian (defined in Eq. (\ref{def_H})). The optimized state $\left| {\psi _* } \right\rangle \equiv  \left| {\psi \left( { - \pi /4,\pi /2} \right)} \right\rangle$ (see Eq. (\ref{trial_state})) (black line) contains a linear combinations of the two eigenstates (red and blue lines) only. (b) The three-qubit NMR quantum simulator consists of a sample of  $^{13}$C-labeled Diethyl-fluoromalonate dissolved in $^{2}$H-labeled chloroform. The nuclear spins (circled) of $^{13}$C and $^1$H are used as the system qubits and that of $^{19}$F is the probe qubit. The parameters of the NMR couplings of this molecule are listed in the table. }\label{Fig1}
\end{figure}

Several theoretical studies \cite{Aspuru2005, Wang2008, Veis2010, Whitfield2011}  along this line of reasoning have been carried out for various molecular structures. Here we performed an experimental realization of this idea with one of the simplest, yet non-trivial, physical systems, namely the Heisenberg spin model in an external field. Our goals for this study are: (i)~to determine the eigenvalues of the ground state, and (ii)~to maximize, or to distill, as much as possible the ground-state from a trial state, which contains a finite ($F=0.5$) ground-state fidelity. For (i), we employed a revised version of the iterative phase estimation procedure to determine the eigenvalues of the Hamiltonian (to the $10^{-5}$ decimal digit). Subsequently, we apply a state-filtering method to extract the ground-state fidelity from the final state to achieved (ii). For this study, we specifically chose three cases corresponding to three different values of external field in the simulation, namely $h=0$, $h=0.75 h_c$, and $h=1.25h_c$, where $h_c$ is the critical value of the external field at which the ground-state and the first excited state cross each other (see Fig.~\ref{Fig1}). This is a singlet-triplet switching, and our experimental simulation captures the change of the ground state around this critical point reliably.

Finally, we note that the approach employed here is different from the method for preparing many-body ground states based on the adiabatic evolution  \cite{Brown2006, Edwards2010, Kim2010, Du2010,Peng2010, Wu2002, Biamonte2011,Chen2011}, where the initial state is usually chosen as the ground state of some simple Hamiltonian, which can be prepared efficiently, instead of the trial states, which aim to capture the essential physics of the exact ground state. The performance (complexity) of the adiabatic approach depends on the energy gap {\it along} the entire evolution path. In our approach, the performance depends on the fidelity of the initial state and the energy gap of the Hamiltonian. Furthermore, in these experiments (except Ref.~\cite{Du2010}), the eigen-energy and the ground state of the Hamiltonian are not usually determined simultaneously, and therefore, cannot be considered as completely solving the ground-state problem \cite{gsp}. In spite of the differences between these two approaches, it is possible that the adiabatic method can be incorporated in our procedure to further enhance the ground-state fidelity of the final state. However, this possible extension is not considered here.

This paper is organized as follows: first we will provide the theoretical background for this experimental work. Then, we define the Hamiltonian to be simulated and the choice and the optimization of the initial state. Next, we outline the experimental procedures, and explain a revised iterative phase estimation algorithm. Finally, the experimental results will be presented and analyzed by a full quantum state tomography. We conclude with a discussion of the results and the sources of errors.


\section{Theoretical background}

The central idea behind this experimental work has a counterpart in the time-domain classical simulation methods \cite{Feit1982}. In the context of quantum computing, the method was introduced by Abrams and Lloyd~\cite{Abrams1999}. Specifically, it was shown that for any quantum state $\left| \psi  \right\rangle  = \sum\nolimits_k {a_k \left| {e_k } \right\rangle }$ which has a finite overlaps $\left| {a_k } \right|^2$ (or fidelity) with the eigenstates $\left| {e_k } \right\rangle$ of a simulated Hamiltonian, $H$, the phase estimation algorithm \cite{Kaye2007} will map, with high probability, the corresponding eigenvalues to the states of an ancilla quantum register,
\begin{equation}\label{PEA_state}
\left| \psi  \right\rangle \left| {000...0} \right\rangle  \to \sum\limits_k {a_k \left| {e_k } \right\rangle \left| {E_k } \right\rangle } \quad.
\end{equation}
Consequently, a {\it projective measurement} on the register qubits will, ideally, collapse the quantum state of the system qubits into one of the eigenstates. By analyzing the measurement outcome, one can determine the ground-state eigenvalue $E_0$, and even project the exact ground state $\left| {e_0 } \right\rangle$.

Given any trial state $\left| {\psi _{T} } \right\rangle$, the performance of the algorithm depends on the overlap $| {a_0}|^2$, which can be maximized using many classical methods, such as using advanced basis sets \cite{Davidson1986}, matrix product states (MPS) representations \cite{Verstraete2006}, or any suitable variational method.


\begin{figure*}[t]
\includegraphics[width= 1.8 \columnwidth]{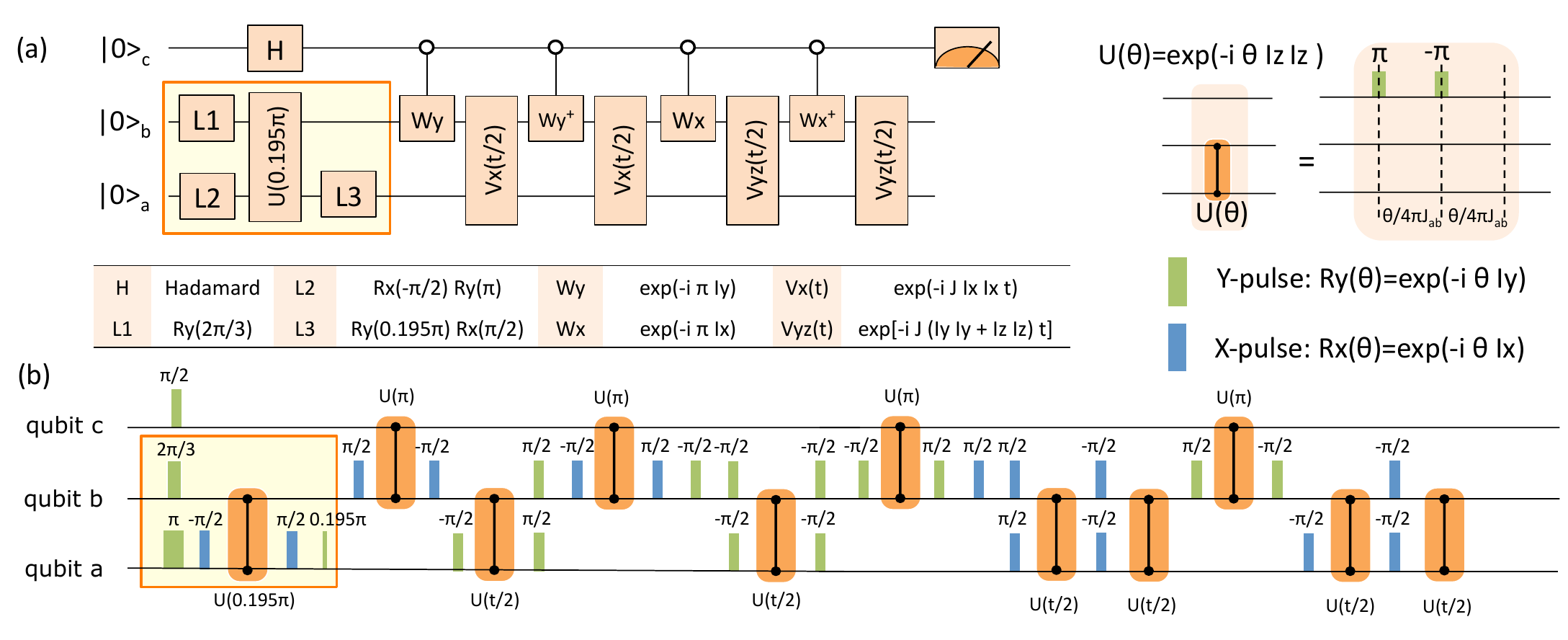}
\caption{(Color online) (a) The quantum circuit diagram for the experiment. The explicit construction of the unitary operators $W$ and $V(t)$ is detailed in the Appendix \cite{appendix}. The boxed quantum gates in (a) (and pulse sequences in (b)) generate the input state $\left| {\psi _* } \right\rangle$, which is an optimized variational state with respect to the Hamiltonian defined in Eq. (\ref{def_H}). (b) The entire pulse sequence corresponds to the quantum circuit diagram for the case of zero external field, $h=0$. The complexity and the lengths of the pulse sequences for the other cases, namely $h=0.75h_c$ and $h=1.25 h_c$, are roughly the same as this pulse sequence.}\label{Fig2}
\end{figure*}

\section{The Hamiltonian and the optimized input state}
The method proposed here can be generalized to apply to more general Hamiltonians, but as a very good example, we will employ the Heisenberg Hamiltonian with an external magnetic field pointing along the $z$-direction:
\begin{equation}\label{def_H}
H = J\left( {I _x^a I _x^b  + I _y^a I _y^b  + I _z^a I _z^b } \right) + h\left( {I _z^a  + I _z^b } \right) \quad,
\end{equation}
where $I _\alpha ^k$= $\frac{1}{2}\sigma _\alpha ^k$, and $\sigma _\alpha ^k$ is one of the Pauli matrices ($\alpha = x,y,z$) acting on the $k=a,b$ spin. On the other hand, in general, there is no restriction to the choice of a trial state, as long as it is not orthogonal to the ground state (in this case, the ground state algorithm necessarily fails). To mimic the behavior of the commonly-employed trial states of more general systems, we require our trial state to satisfy the following conditions: (a) that it contains one or more parameters which can be adjusted to minimize the energy $\left\langle H \right\rangle$, and that this procedure usually does not lead to the exact ground state, and (b) that it may capture only part of the vector space spanned by the eigenstates of the Hamiltonian $H$. One possible choice that fulfills the above criteria is the following variational state which contains two adjustable parameters, $\theta$ and $\phi$,
\begin{equation}\label{trial_state}
\left| {\psi \left( {\theta ,\varphi } \right)} \right\rangle  = \frac{1}{{\sqrt 2 }}\left| \theta  \right\rangle  + \frac{1}{{\sqrt 2 }}\left| \varphi  \right\rangle \quad .
\end{equation}
Here, $\left| \theta  \right\rangle  \equiv \cos \theta \left| {10} \right\rangle  + \sin \theta \left| {01} \right\rangle$ and $\left| \varphi  \right\rangle  \equiv \cos \varphi \left| {00} \right\rangle  + \sin \varphi \left| {11} \right\rangle$. In general, the optimized states for each given pair of $(J,h)$ are not necessarily the same. However, in our case, we found that the optimized state $\left| {\psi _* } \right\rangle \equiv  \left| {\psi \left( { - \pi /4,\pi /2} \right)} \right\rangle$ can minimize the energy for all values of $h$ and $J>0$. Moreover, it turns out that this optimized state captured two out of the four eigen-energies (see  Fig.~\ref{Fig2}a) only; therefore, a single probe qubit is sufficient to resolve them (for more general cases, see the Appendix \cite{appendix}). We note that the fidelity $F$ (cf. Eq.~(\ref{def_fidelity})) of the state $\left| {\psi _* } \right\rangle$ with the exact ground state $\left| {e_0 } \right\rangle$ is exactly 50\%.

\section{Outline of the method}
This algorithm starts with a set of system qubits initialized in the state $\left| \psi_*  \right\rangle  = \sum\nolimits_k {a_k \left| {e_k } \right\rangle }$ and a single ``probe" qubit in the $\left( {\left| 0 \right\rangle  + \left| 1 \right\rangle } \right)/\sqrt 2$ state. For different times $t$, a controlled $U(t)$ gate, where $U\left( t \right) \equiv e^{ - i H t}$ ($\hbar=1$), is then applied, resulting in the following state: $\left( {1/\sqrt 2 } \right)\sum\nolimits_k {a_k \left( {\left| 0 \right\rangle  + e^{ - i\omega _k t} \left| 1 \right\rangle } \right)\left| {e_k } \right\rangle }$, where $\omega _k  \equiv E_k $. The reduced density matrix of the probe qubit,
\begin{equation}\label{probe_qubit}
\rho_{ probe} (t)  = \frac{1}{2}\left( {\begin{array}{*{20}c}
   1 & {\sum\nolimits_k {\left| {a_k } \right|^2 e^{i\omega _k t} } }  \\
   {\sum\nolimits_k {\left| {a_k } \right|^2 e^{ - i\omega _k t} } } & 1  \\
\end{array}} \right) \, ,
\end{equation}
contains the information about the eigenvalues in its off-diagonal matrix elements, which can be measured efficiently in an NMR setup (see Appendix \cite{appendix}). A classical Fourier analysis on the off-diagonal matrix elements at different times yields both the eigenvalues $\omega_k$ and the overlaps ${\left| {a_k } \right|^2 }$. To obtain the value of $\omega_k$ with high accuracy, a long time evolution of the simulated quantum state is usually needed. However, for Hamiltonians with certain symmetries,
we can perform a simplified version of the iterative phase estimation algorithm (IPEA), which is similar but not identical to the ones performed previously in Ref. \cite{Lanyon2010, Du2010}. We will explain the details of this IPEA in Section \ref{sec:IPEA}.

\begin{figure}[tb]
\includegraphics[width=  0.9 \columnwidth]{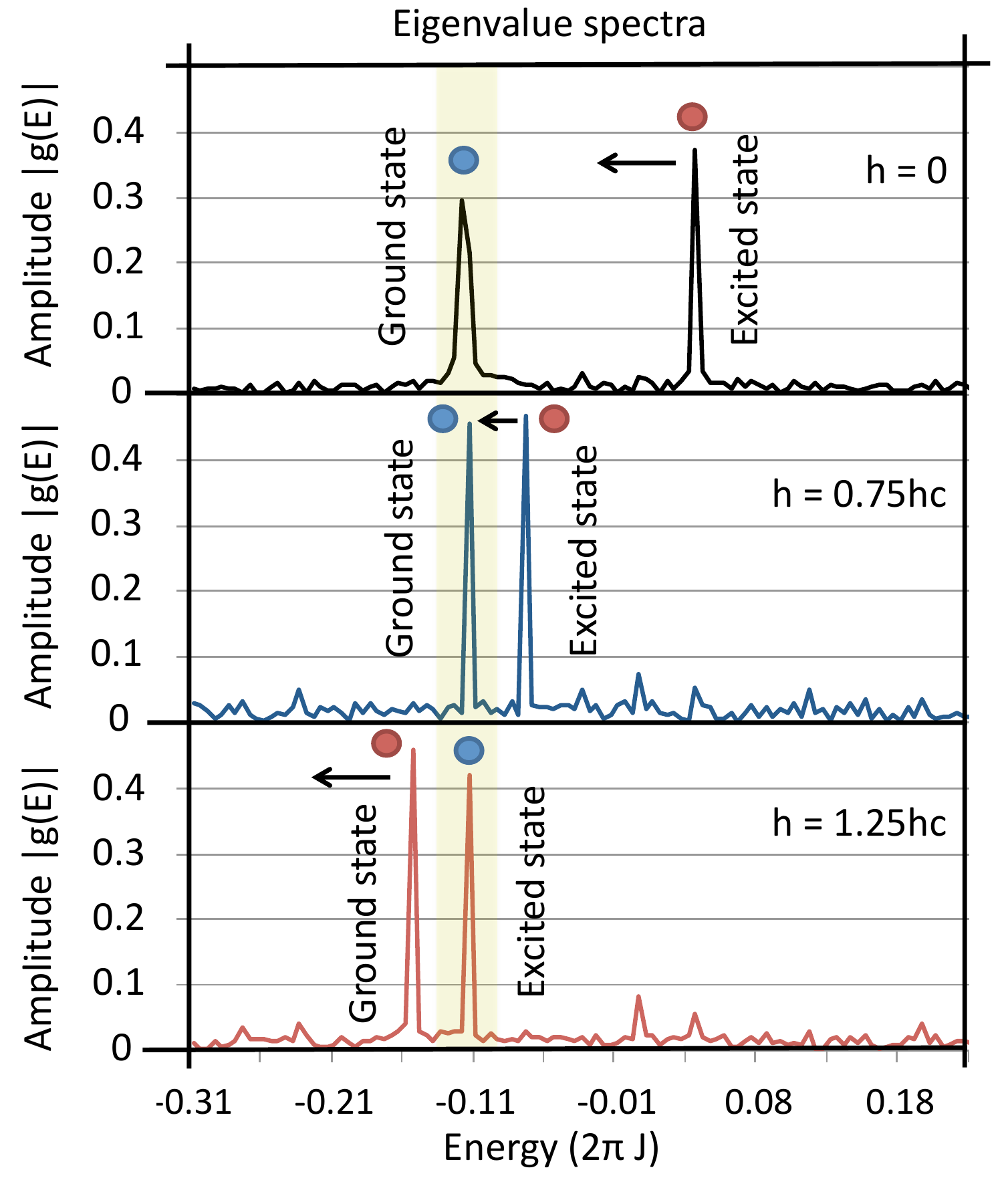}
\caption{(Color online)  The absolute amplitude of the eigenvalue spectra $g(E)$ of the Heisenberg Hamiltonian as defined in Eq.~(\ref{def_H}). These are obtained for three different values of the external magnetic field $h$, namely $h=0$, $h=0.75 h_c$, and $h=1.25 h_c$, where $h_c$ is the critical field at which the ground state becomes degenerate. The shaded region highlights the location the the singlet state depicted in Fig.~\ref{Fig1}. The arrows indicate the direction of the peak shift when the simulated external field $h$ is increased. The blue and red dots indicate the quantum states represented by the peak signals (cf. Fig.~\ref{Fig1}).}\label{Fig3}
\end{figure}

Once the ground state eigenvalue $E_0$ of the Hamiltonian $H$ is determined, one can, for example, employ the state-filtering method \cite{Poulin2009} to isolate the corresponding state from the rest. Following, measurement of any observable, and even quantum state tomography, can be performed for the resulting ground state. With the complete knowledge of the eigenstates and the eigenvalues, we can in principle obtain all accessible information about the ground state properties; therefore, this procedure solves the ground-state problem when trial wave functions are available.

\section{Experimental procedure}
The experiments were carried out at room temperature on a Bruker AV-400 spectrometer. The sample we used is the $^{13}$C-labeled Diethyl-fluoromalonate dissolved in $^{2}$H-labeled chloroform. This system is a three-qubit quantum simulator using the nuclear spins of $^{13}$C and $^1$H as the system qubits to simulate the Heisenberg spins, and the $^{19}$F as the probe qubit in the phase estimation algorithm (see Fig.~\ref{Fig1}b). The internal Hamiltonian $H_{NMR}$ of this system can be described by the following:
\begin{equation}\label{Hamiltonian}
H_{NMR} = \sum\limits_{j \in \{a,b,c\}}  2 \pi  {\nu _j } I_z^j  +   \sum\limits_{j < k \in \{a,b,c\} } 2 {\pi} J_{jk} I_z^j I_z^k \quad ,
\end{equation}
where $\nu_j$ is the resonance frequency of the \emph{j}th spin and $\emph{J}_{jk}$ is the scalar coupling strength between spins \emph{j} and \emph{k}, with $\emph{J}_{ab}=160.7$~Hz, $\emph{J}_{bc}=-194.4$~Hz, and $\emph{J}_{ac}=47.6$~Hz. The relaxation time $T_1$ and dephasing time $T_2$ for each of the three nuclear spins are tabulated in Fig.~\ref{Fig1}a.

The experimental procedure consists of three main parts: I. State initialization (preparing the system qubits as $\left| {\psi _* } \right\rangle$, probe qubit as  $\left| {0 } \right\rangle$), II. Eigenvalue measurement by iterative phase estimation, and III. Quantum state tomography. The state initialization part is rather standard and we leave the details of it to the Appendix \cite{appendix}. Part II is implemented with a quantum circuit as depicted in Fig.~\ref{Fig2} (see the Appendix \cite{appendix} for the detailed circuit construction). The probe qubit is measured at the end of the circuit (see also Eq. (\ref{probe_qubit})).

The resulting Fourier spectra for various cases are shown in Fig.~\ref{Fig3}. The positions of the peaks indicate the eigenvalue of the Hamiltonian $H$. Although the peaks look sharp, the errors are in fact about $22\%$. However, we are able to reduce the errors to less than $0.003\%$ (see Fig.~\ref{Fig4}) by five steps of the iterative phase estimation algorithm which is described below.

\section{Iterative phase estimation algorithm (IPEA)}\label{sec:IPEA}
\begin{figure*}[t]
\includegraphics[width=  1.8 \columnwidth]{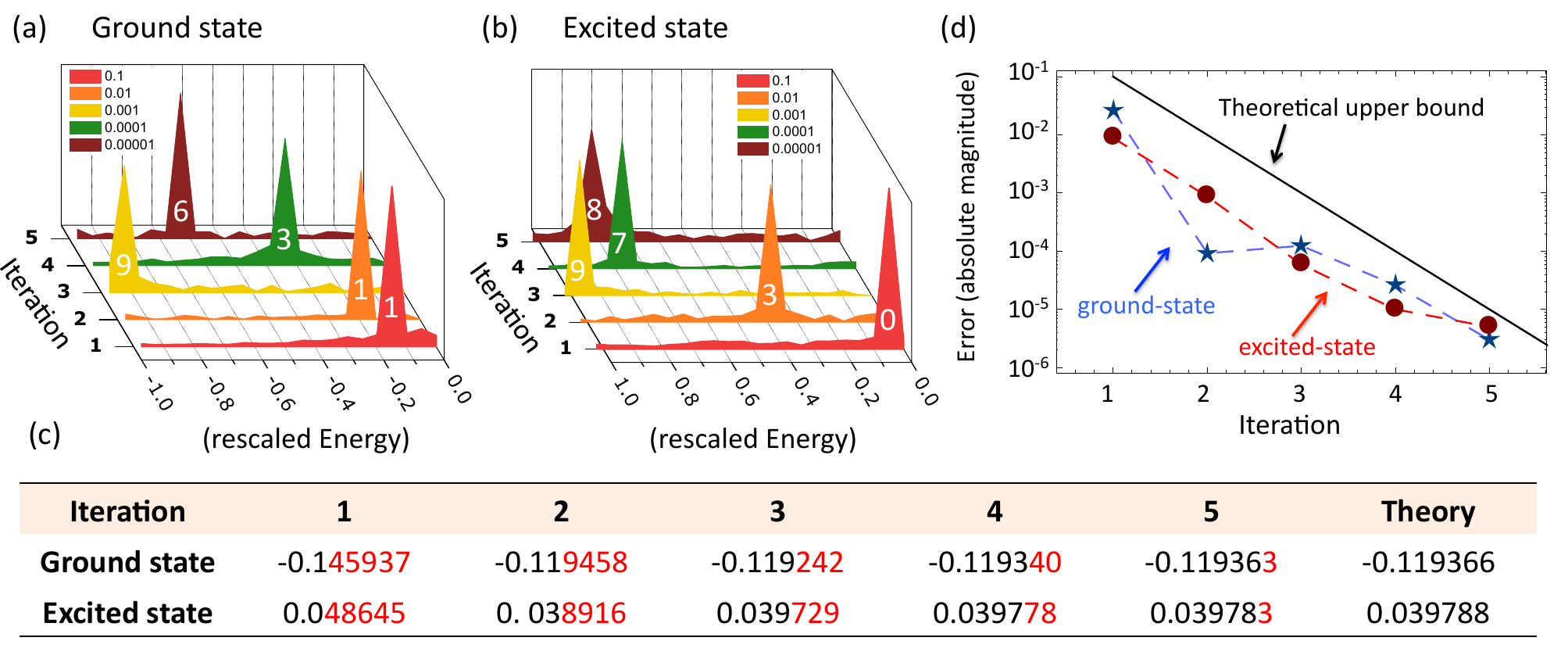}
\caption{(Color online) Experimental results of the iterative phase estimation algorithm (IPEA) for improving the accuracy of the measured ground-state energy. (a) and (b) There are five iterations performed; each of them improves one digit of accuracy in the eigenvalues. For example, consider the ground state, after the first iteration (red curve), the peak lies between $-0.1$ and $-0.2$; this means that the first digit of eigenvalue should be $-0.1$. After five iterations, the value of ground-state energy is determined to be $-0.11936(3)$,  with a precision of $10^{-5}$ in units of $2\pi J$. (c) A table listing the improvement of the numerical values (digits in red represent uncertainty). (d) Graphical visualization of the results in (c). The theoretical curve results from the improvement of the precision by a factor of 1/10 for each iteration.}\label{Fig4}
\end{figure*}
To improve the resolution of the energy eigenvalues, the information stemming from long time evolution of the simulated state is needed \cite{Brown2006}. Fortunately, the required resources can be significantly reduced by the IPEA approach. This is due to the symmetry of the Hamiltonian: since all the terms in the Hamiltonian (Eq. (\ref{def_H})) commute with each other, they can be simulated individually, i.e.,
\begin{equation}
e^{ - iHt}  = e^{ - iJI _x^a I _x^b t} e^{ - iJI _y^a I _y^b t} e^{ - iJI _z^a I _z^b t} e^{ - ih\left( {I _z^a  + I _z^b } \right)t}
\end{equation}
for all times $t$. The last term $e^{ - ih\left( {I _z^a  + I _z^b } \right)t}$ corresponds to two separate local rotations, whose implementation is straight-forward (see the Appendix \cite{appendix}). The other terms $e^{ - i J I _\alpha ^a I _\alpha ^b t}$ are equivalent up to some local unitary rotations, and their eigenvalue spectra of $I _\alpha ^a I _\alpha ^b$, which are $4$ and $-4$, are the same; the eigenvalues are symmetrical about zero. This means that, in order to simulate each term for a time interval $t$, we can always find a shorter time $\tau$ such that $e^{ - iJI _\alpha ^a I _\alpha ^b \tau }  {=} e^{ - iJI _\alpha ^a I _\alpha ^b t} $, where $t = 8n\pi /J + \tau$ for some non-negative integer $n$ which is determined by the condition: $0 \le J\tau  \le 8\pi $.

\begin{figure}[t]
\includegraphics[width=   0.9 \columnwidth]{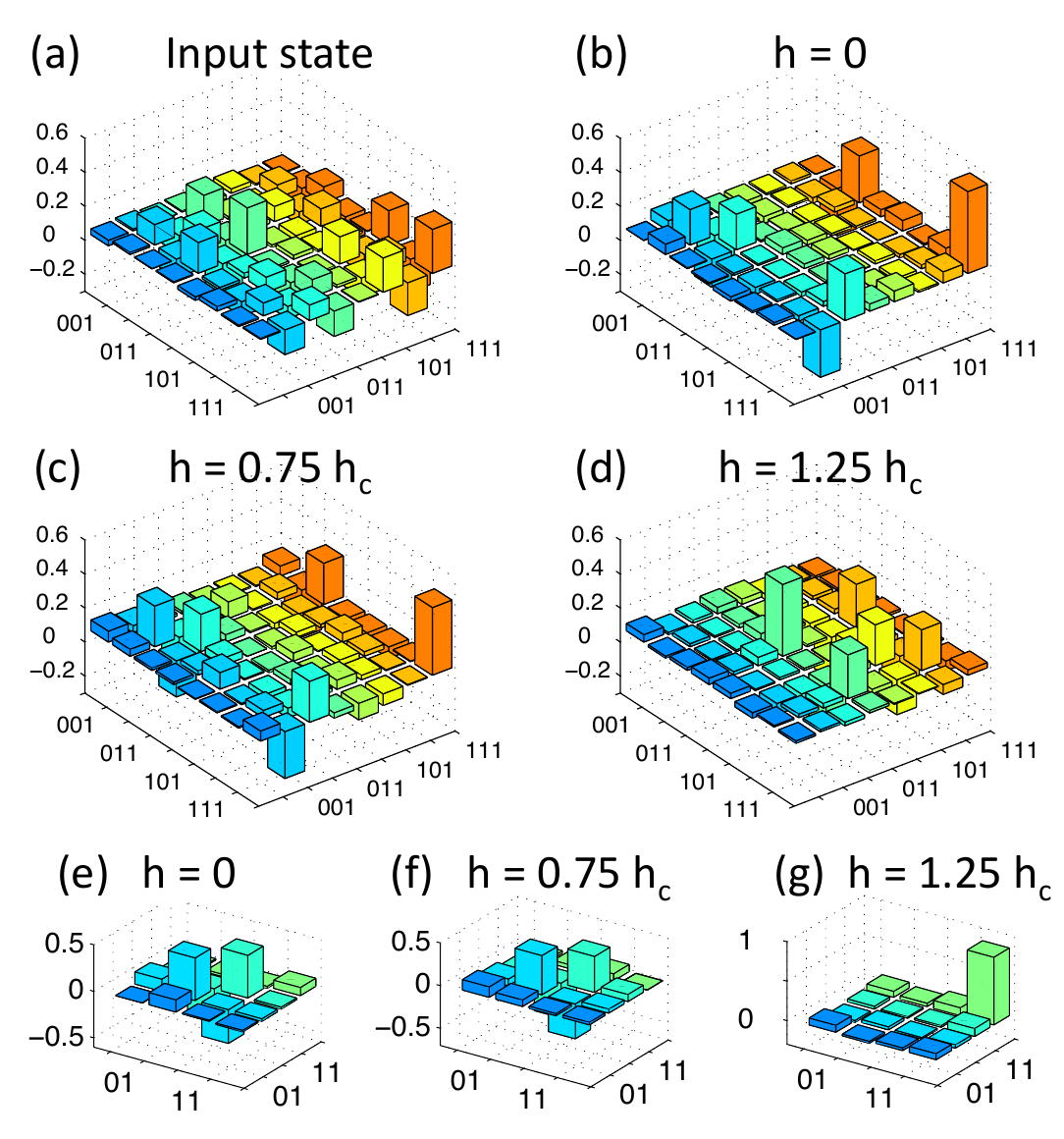}
\caption{(Color online) Experimental results from the quantum process tomography procedure (real parts are shown, imaginary parts shown in the Appendix \cite{appendix}). (a) the initial state $\left\vert \psi_{*} \right\rangle $. (b),(c), and (d) Three final states (Eq. (\ref{fin_state})) for the cases, respectively, $h=0$, $h=0.75 h_c$, and $h=1.25 h_c$. (e),(f), and~(g) The first $4\times4$ section of each density matrix above (after re-normalization), in the subspace where the probe qubit is projected to the $\left| 0 \right\rangle$ state.}\label{Fig5}
\end{figure}

Now, denote the eigenvalue, $\omega _k \equiv 2\pi J  \times 0.x_1 x_2 x_3 ... $, by a string of decimal digits $ \{ x_1, x_2, x_3 ... \}$. The first digit $x_1$ can be determined by a short time evolution by a probe qubit described in Eq. (\ref{probe_qubit}). Once $x_1$ is known, the second digit $x_2$ can be iteratively determined by simulating the evolution for ten times longer than the previous ones:
\begin{equation}
10 \times \omega _k t = 2\pi  Jt \times x_1.0 + 2\pi Jt \times 0.x_2 x_3 ...
\end{equation}
Note that the first term on the right hand side is known. The second term is now amplified, and can be resolved by the probe qubit. This means that the eigenvalue $\omega_k$ can then be determined to two digits of precision. By repeating this scheme iteratively for $x_3$ and so on, the eigenvalue $\omega_k$ can be determined subsequently for one digit after the other (cf. Fig.~\ref{Fig4}). The accuracy of the eigenvalues is improved from about $22\%$ to about $0.003\%$. We note that in the IPEA performed in Refs. \cite{Lanyon2010, Du2010}, the final unitary matrices are decomposed directly for each value of $t$. Therefore, one can in principle determine the eigenvalues to any arbitrary accuracy. However, the resources required for decomposing the unitary matrices grow exponentially with the system size; the methods implemented there are certainly unrealistic for larger systems. Here, we exploited the symmetry of the Hamiltonian, and simulate the time evolution without performing the decomposition of the unitary matrices. The accuracy of the IPEA is limited by some natural constraints. The details about the limitation of this method are discussed in the Appendix~\cite{appendix}.

\section{Results and Discussion}
\begin{figure}[t]
\includegraphics[width=  0.9 \columnwidth]{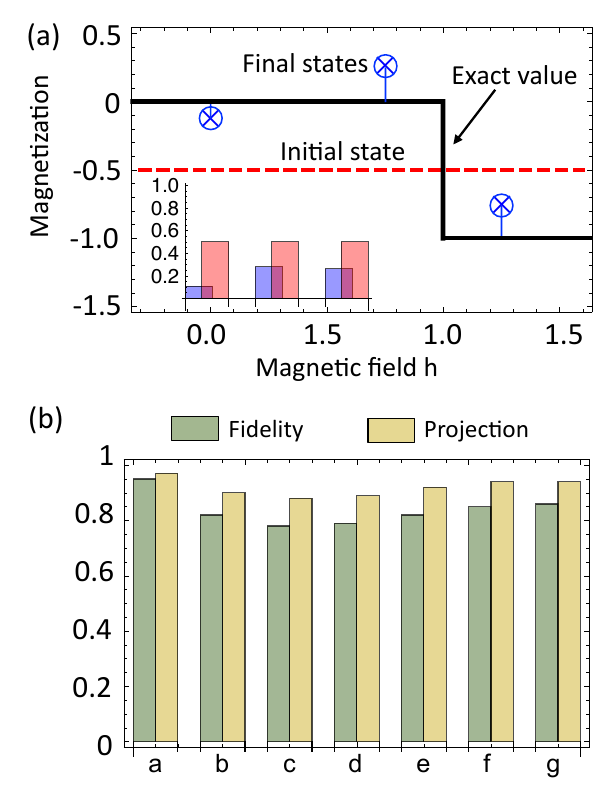}
\caption{(Color online) Results from the quantum state tomography. (a) Magnetization $\langle I_z^1 \rangle + \langle I_z^2 \rangle$ for the initial states (red dotted line) and the final states (crossed circles) for $h=0$, $h=0.75h_c$, and $h=1.25 h_c$. The inset shows the absolute errors of the initial state (red bars), and the three experimental values (blue bars). (b) The ground state fidelity (green) and projection (yellow) for the experimentally determined states (a)-(g) in Fig.~\ref{Fig5}. The fidelity of the initial state (blue) is included for comparison.}\label{Fig6}
\end{figure}

\begin{figure}[t]
\includegraphics[width=   0.9 \columnwidth]{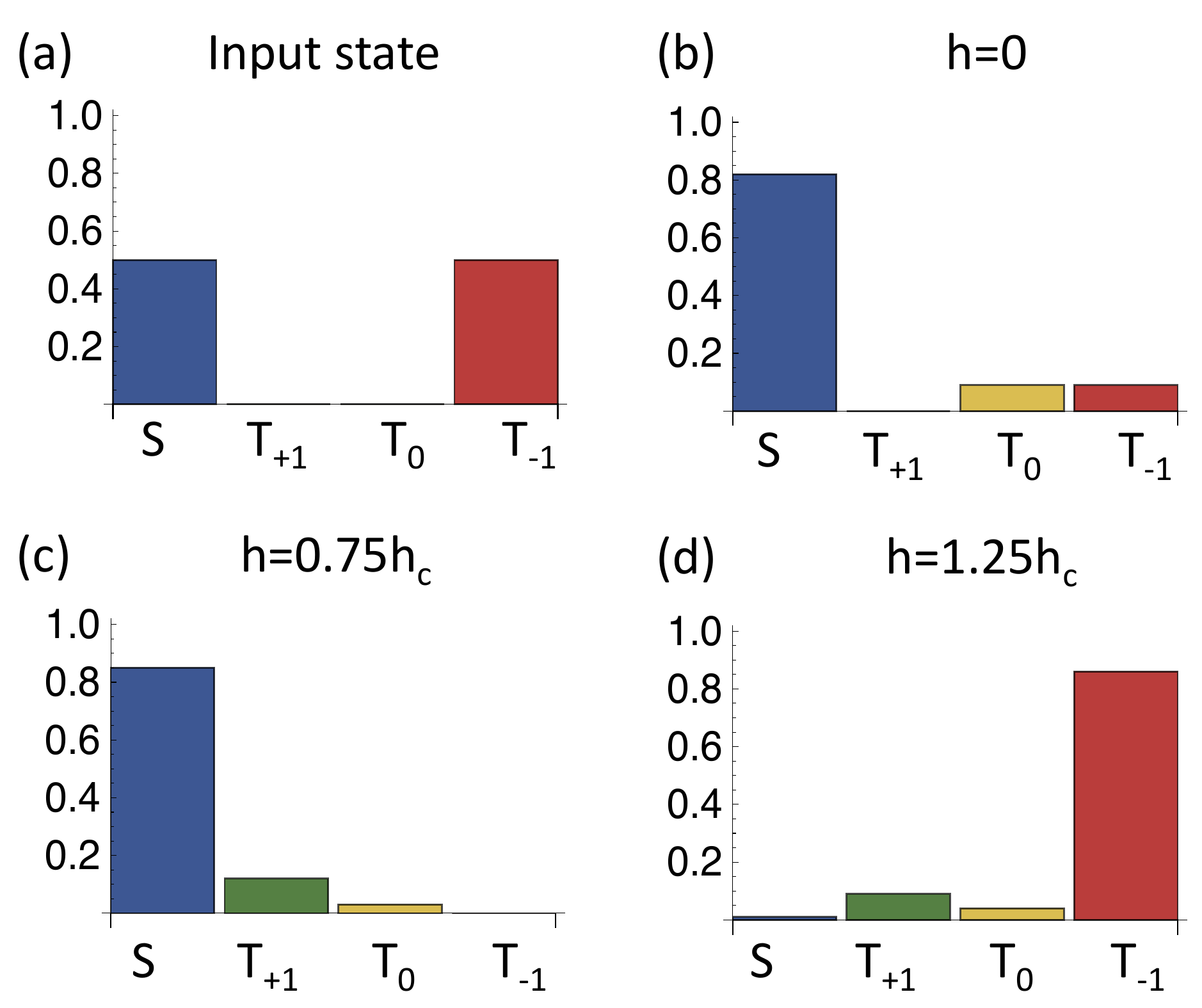}
\caption{(Color online) Spectral decomposition of the final states. Panels (a)-(d) show the weights (probability) of the eigenstates, $\rm S=(| 01\rangle- |10\rangle)/\sqrt 2 $, $\rm T_{+1}=| 00\rangle$, $\rm T=(| 01\rangle +|10\rangle)/\sqrt 2$, and $\rm T_{-1}=|11\rangle$, of the Heisenberg Hamiltonian in the initial and final states.}\label{Fig7}
\end{figure}

Once the two eigenvalues ($E_0$ and $E_1$) are accurately determined by the IPEA, we can identify the eigenvectors (ground state $\left| {e_0 } \right\rangle$ and excited state $\left| {e_1 } \right\rangle$) by the same quantum circuit as shown in Fig.~\ref{Fig2}a. The difference is that, the time $\tau$, in the controlled rotation $U\left( \tau  \right) \equiv e^{ - iH\tau }$ is chosen to be $\tau  = \pi /\left( {E_1  - E_0 } \right)$. This allows us to obtain the following state,
\begin{equation}\label{fin_state}
\frac{1}{{\sqrt 2 }}\left( {\left| {e_0 } \right\rangle \left| 0 \right\rangle  - \left| {e_1 } \right\rangle \left| 1 \right\rangle } \right) \quad .
\end{equation}
This state is very similar to the one discussed in Eq.~(\ref{PEA_state}). The important point is that, now each eigenstate is tagged by the two orthogonal states of the ancilla qubit, and can be determined separately, e.g. through quantum state tomography.

To obtain the state in Eq (\ref{fin_state}), starting from the product state $\left| \psi_*  \right\rangle \left| 0 \right\rangle $, we first prepared the probe state as a superposition state with a phase $e^{i E_0 \tau }$ ``preloaded" in it, i.e., $\left( {\left| 0 \right\rangle  + e^{i E_0 \tau } \left| 1 \right\rangle } \right)/\sqrt 2$. Next, after applying the controlled-$U(\tau)$ to the trial state $\left| \psi_*  \right\rangle  = {a_0 \left| {e_0} \right\rangle } + {a_1 \left| {e_1} \right\rangle }$, we have,
\begin{equation}
\frac{1}{{\sqrt 2 }}a_0 \left| {e_0 } \right\rangle \left( {\left| 0 \right\rangle  + \left| 1 \right\rangle } \right) + \frac{1}{{\sqrt 2 }}a_1 \left| {e_1 } \right\rangle \left( {\left| 0 \right\rangle  + e^{i\pi } \left| 1 \right\rangle } \right) \quad.
\end{equation}
Subsequently, we apply a single-qubit rotation gate $R_y^c( - \pi/2)$, which maps $  {(\left| 0 \right\rangle  + \left| 1 \right\rangle) /\sqrt 2 }
 \to \left| 0 \right\rangle $ and $ \left( {\left| 0 \right\rangle  - \left| 1 \right\rangle } \right)/\sqrt 2 \to - \left| 1 \right\rangle $, we then obtain the final state in Eq.~(\ref{fin_state}).

Finally, the standard procedure of quantum state tomography \cite{Leskowitz2004} was performed on the final states (Eq. (\ref{fin_state})) for the cases $h=0$, $h=0.75h_c$, and $h=1.25 h_c$, shown respectively in Fig.~\ref{Fig5} (b)-(d). The corresponding results of the ground state (i.e. the $\left| {e_0 } \right\rangle$ part in Eq. (\ref{fin_state})) are shown in Fig.~\ref{Fig5}~{(e){-}(g)}. These density matrices allow us to obtain all information about the experimentally determined ground states. Fig.~\ref{Fig6}a shows the improvement of the magnetization $M$ of the final states, as compared with the initial state. The inset figure shows that the magnitude of the deviations (blue bars) from the theoretical values are always smaller then that (red bars) of the trial state.

The quality of the final state $\rho_{exp}$ in the experiment is quantified by the fidelity $F=\left\langle e_{0} \right\vert\rho_{exp}\left\vert e_0 \right\rangle $ (cf. Eq. (\ref{def_fidelity})), and the projection \cite{Fortunato2002} $P=F/\sqrt{Q}$, where $Q = {\rm Tr}(\rho _{exp}^2 )$ is the purity of $\rho_{exp}$. The results are shown in Fig.~\ref{Fig6}b. Note that the reduced density matrices (e),(f),(g) have better fidelities than that of the original density matrices (b),(c),(d). In Fig.~\ref{Fig7}, the weights (probabilities) of the eigenstates of $H$ in the final states are shown. Note that, as mentioned above, the trial state captures only two eigenstates. Due to experimental errors, other eigenstates also showed up in the spectral decomposition. This contributes to the deviation of the magnetization ($M=0$ for the singlet state) as well. Note that the singlet-triplet switching (cf. Fig.~\ref{Fig1}), i.e., from Fig.~\ref{Fig7}c to \ref{Fig7}d, is reliably captured.

In this experiment, we are able to determine the eigenvalues to a very high accuracy, using the iterative phase estimation algorithm (IPEA). The major source of errors (about $10\%$ of the fidelity) of the experiment comes from the second step of the procedure where the overall pulse sequence to construct the final state Eq. (\ref{fin_state}) is lengthy, and therefore is dominantly a $T_2$ error; the time spent for this operation is about 1/10 of $T_2$ (see the Appendix~\cite{appendix}). Additionally, other errors come from the measurement (tomography) errors, and the inhomogeneity in the RF pulses and the external magnetic field. If these factors can be overcome, a further increase of fidelity is possible by using the final state of this experiment as the input state for another iteration of the similar distillation procedure (see the Appendix~\cite{appendix} for details).

\section{Conclusion}

We have experimentally demonstrated a method to solve the quantum ground-state problem using an NMR setup. This is achieved by distilling the exact ground state from an input state, which has 50\% overlap with the ground state. The eigenvalues were determined to a precision of the $10^{-5}$ decimal digit, after five iterations of the phase estimation procedure. Then, the final states are distilled to high values of fidelity.  The method we developed in this experiment is scalable to more general Hamiltonians, and not limited to NMR systems. This result confirms that variational methods developed for classical computing could be a good starting point for quantum computers, opening more possibilities for the purposes of quantum computation and simulation.



\begin{acknowledgments}
We are grateful to the following funding sources: Croucher Foundation for M.H.Y; DARPA under the Young Faculty Award N66001-09-1-2101-DOD35CAP, the Camille and Henry Dreyfus Foundation, and the Sloan Foundation;   NSF Center for Quantum Information and Computation for Chemistry, Award number CHE-1037992 for A.A.G. This work is also supported by the National Nature Science Foundation of China, the CAS, and the National Fundamental Research Program 2007CB925200.
\end{acknowledgments}

\bibliographystyle{apsrev}

\newpage
\begin{widetext}
\center
{\bf Supplementary materials: Experimental Implementation of Quantum Ground-State Distillation}
\medskip
\\ Zhaokai Li, Man-Hong Yung, Hongwei Chen, Dawei Lu, \\ James D. Whitfield, Xinhua Peng, Al\'{a}n Aspuru-Guzik, and Jiangfeng Du
\bigskip
\end{widetext}

\appendix

\tableofcontents


\bigskip

\section{State initialization}\label{app_state_in}

\begin{figure}[h]
\includegraphics[width= 0.9 \columnwidth]{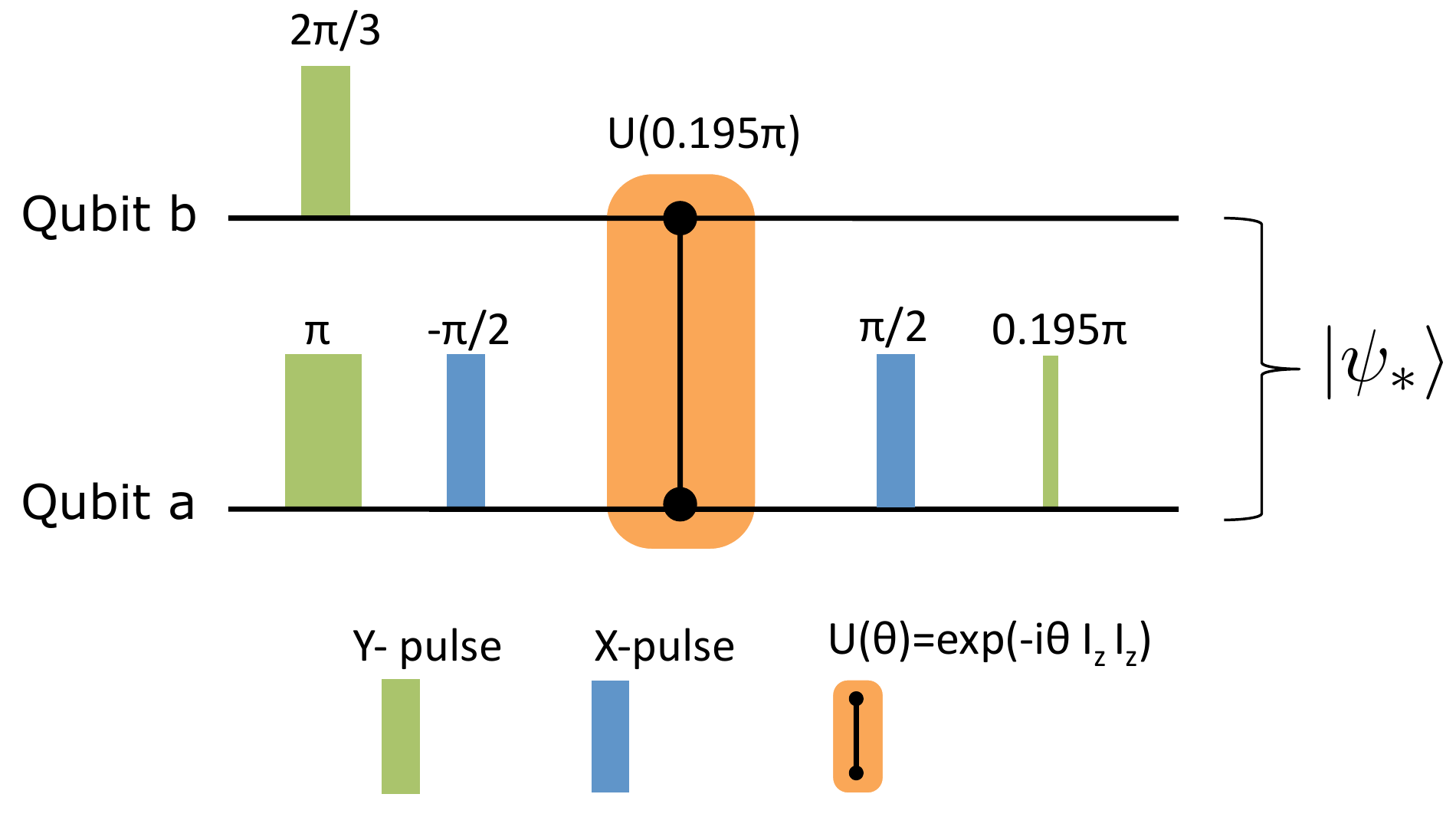}
\caption{(Color online) Pulse sequence for generating the input state $\left\vert \psi_{*} \right\rangle$.}\label{apdx_pulse_1}
\end{figure}

In this experiment, we used a sample of the $^{13}$C-labeled Diethyl-fluoromalonate dissolved in the $^2$H-labeled chloroform as a three-qubit computer, where the nuclear spins of the $^{13}$C and the $^1$H were used as the system qubits, and that of the $^{19}$F was used as the probe qubit. The structure of the molecule is shown in Fig.~\ref{Fig1}a of the main text, and the physical properties are listed in the table of Fig.~\ref{Fig1}b.

Starting from the thermal equilibrium state, we first created the pseudo-pure state (PPS)
\begin{equation}
\rho_{000}=(1-\epsilon)\mathbb{{I}}/8+\epsilon \left\vert 000 \right\rangle \left\langle000\right\vert
\end{equation}
using the standard spatial average technique. Here, $\epsilon {\approx} 10^{-5}$ quantifies the strength of the polarization of the system, and ${\mathbb{{I}}}$ is the $8\times 8$ identity matrix. Next, we prepared the probe qubit to the state $\frac{1}{\sqrt{2}}(\left\vert 0 \right\rangle +\left\vert 1 \right\rangle)$ by a pseudo-Hadamard gate $R_y^c(\pi/2)$, where,
\begin{equation}
R_\alpha^j  \left( \theta  \right) \equiv e^{ - i\theta I _\alpha^j} \quad .
\end{equation}
Here, $\alpha = x,y,z$, is a rotation operation applied to the qubit~$j$.

Finally, the system qubits are prepared to the initial state,
\begin{eqnarray}\label{initial}
\left\vert \psi_{*} \right\rangle = \frac{1}{2}(\left\vert 01 \right\rangle-\left\vert 10 \right\rangle)+\frac{1}{\sqrt{2}}\left\vert 11 \right\rangle \quad,
\end{eqnarray}
by applying two single-qubit rotations and one controlled-rotation. The pulse sequence employed follows:
\begin{eqnarray}
R_y^a(\pi) \, \rightarrow \, R_y^b({\textstyle{2 \pi \over 3}}) \, &\rightarrow& \, R_x^a({\textstyle{- \pi \over 2}})  \, \rightarrow  \, U^{ab}(0.195\pi)\, \nonumber \\
&\rightarrow& \, R_x^a({\textstyle{\pi \over 2}}) \, \rightarrow \, R_y^a(0.195\pi),
\end{eqnarray}
where the unitary evolution,
\begin{equation}
U^{jk}(\theta) \equiv e^{-i\theta I^j_z I^k_z}
\end{equation}
is generated from the natural evolution between qubit $j$ and $k$.

\section{Quantum circuit diagram for simulating the controlled-$U(t)$}

\begin{figure}[h]
\includegraphics[width=0.9 \columnwidth]{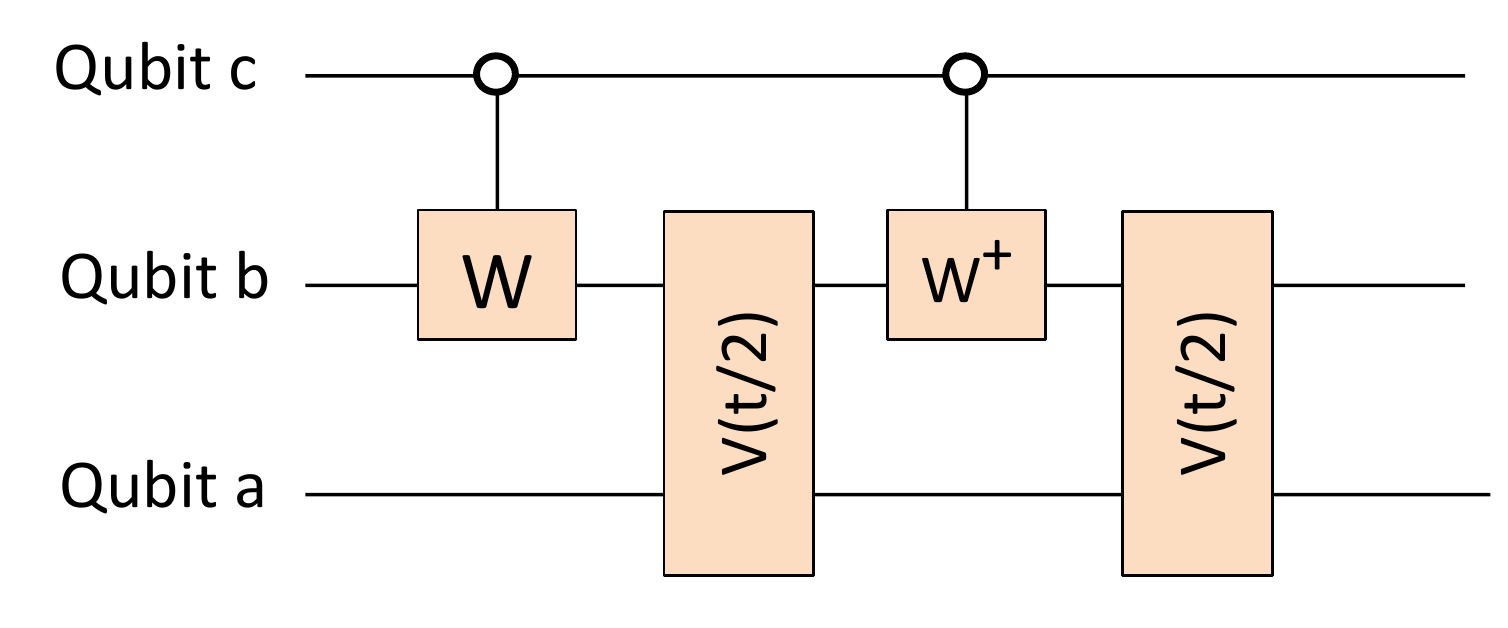}
\caption{(Color online) Quantum circuit diagram for simulating controlled-$V_x$ and controlled-$V_{yz}$.}\label{apdx_fig_U}
\end{figure}

\begin{figure}[h]
\includegraphics[width=0.9 \columnwidth]{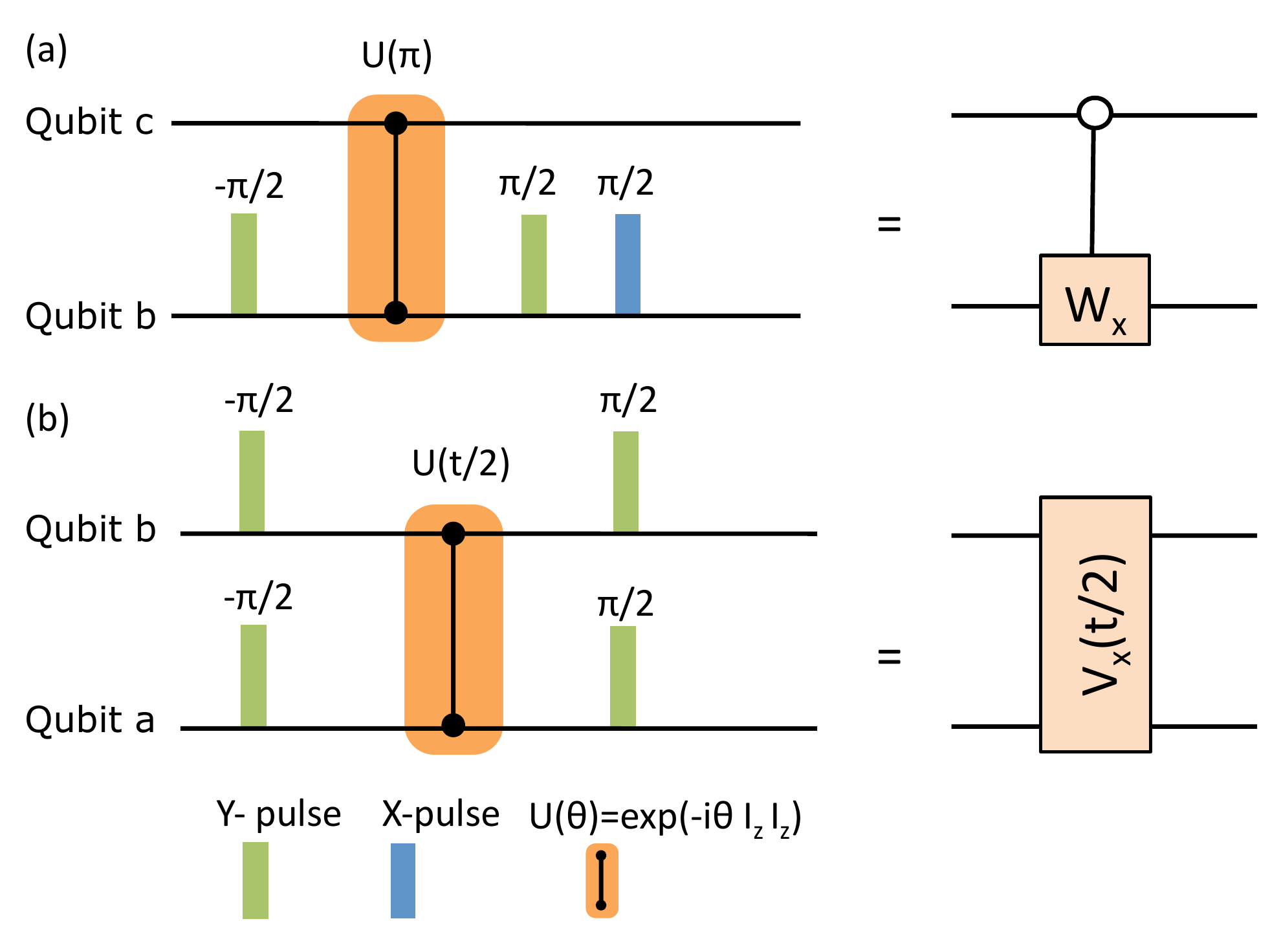}
\caption{(Color online) Pulse sequences for simulating {controlled{-}$e^{ - i\pi I _x^b}$} and $V_x (t/2)$.}\label{apdx_pulse_2}
\end{figure}

The controlled-$U(t)$ in the phase estimation algorithm (see Fig.~\ref{Fig2}a) is implemented in the following way: since all the terms in the the Heisenberg Hamiltonian,
\begin{equation}\label{apdx_def_H}
H = J \left( {I _x^a I _x^b  + I _y^a I _y^b  + I _z^a I _z^b } \right) + h\left( {I _z^a  + I _z^b } \right) \quad,
\end{equation}
commute with each other, we decompose the time evolution operator $T(t) \equiv e^{ - iHt}$ into three parts:
\begin{equation}
T\left( t \right) = V_x \left( t \right)V_{yz} \left( t \right) L_z \left( t \right) \quad,
\end{equation}
where
\begin{eqnarray}
V_x (t) &\equiv& e^{ - iJI _x^a I _x^b t}, \\
V_{yz} \left( t \right) &\equiv& e^{ - iJ\left( {I _y^a I _y^b  + I _z^a I _z^b } \right)t}, \\
L_z \left( t \right) &\equiv& e^{ - ih\left( {I _z^a  + I _z^b } \right)t}.
\end{eqnarray}

 The quantum circuit diagram for simulating the operations controlled-$V_x$ and controlled-$V_{yz}$ is shown in Fig.~\ref{apdx_fig_U}. To simulate controlled-$V_x(t)$, we set,
 \begin{equation}
 V(t/2)=V_x(t/2) \quad {\rm and} \quad W_y=e^{ - i\pi I _y} \quad .
\end{equation}
 (alternatively, $I_z$); to simulate controlled-$V_{yz}(t)$, we set
 \begin{equation}
 V(t/2)=V_{yz}(t/2) \quad {\rm and} \quad W_x=e^{ - i\pi I _x} \quad.
\end{equation}
 Note that the control is ``on" when the probe qubit is in the $\left| 0 \right\rangle$ state. In this case, the first three quantum gates cancel the last gate $V(t/2)$, making it effectively an identity gate. When the controlling qubit is in the ``off" state, this circuit executes two $V(t/2)$ gates.

The pulse sequences for generating the {controlled{-}$e^{ - i\pi I _x^b}$} gate are,
\begin{equation*}
 R_y^b({\textstyle{-\pi \over 2}})\rightarrow U^{bc}(0.5) \rightarrow R_y^b({\textstyle{\pi \over 2}}) \rightarrow R_x^b({\textstyle{\pi \over 2}})\quad,
\end{equation*}
and the pulse sequences of the $V_x (t/2)$ gate is:
\begin{equation*}
 \{ R_y^{a} ({\textstyle{-\pi \over 2}}), R_y^{b} ({\textstyle{-\pi \over 2}}) \} \rightarrow U^{ab}(J t/2) \rightarrow  \{ R_y^{a} ({\textstyle{\pi \over 2}}),  R_y^{b} ({\textstyle{\pi \over 2}})\} \quad.
\end{equation*}
The corresponding diagrams of the pulse sequence are shown in Fig.~\ref{apdx_pulse_2}.

\section{Measurement of the probe qubit}
\begin{figure}[t]
\includegraphics[width=0.8  \columnwidth]{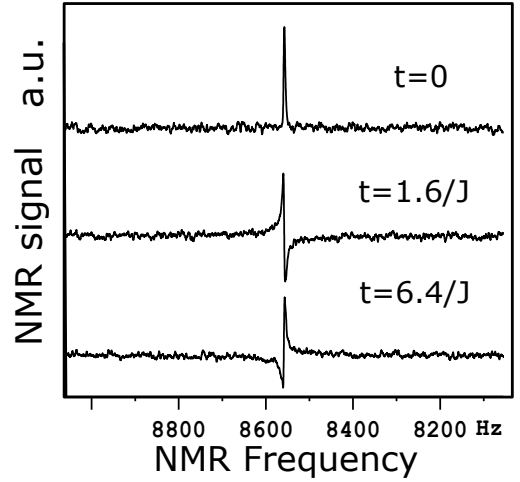}
\caption{The signals of the experimental spectra for the case $h=0$, where $t=0,1.6/J$ and $6.4/J$.}\label{apdx_signal}
\end{figure}
Here we explain the measurement method of the NMR signal of the probe qubit (see Eq. (\ref{probe_qubit})). Denote the off-diagonal elements of $\rho_{probe}(t)$ as,
\begin{equation}
\left| {M_t } \right|e^{i\phi _t }  \equiv \sum\limits_k {\left| {a_k } \right|} ^2 e^{i\omega _k t} \quad.
\end{equation}
The phase shift $\phi_t$ can be obtained by using the method of quadrature detection which serves as a phase detector. By measuring the integrate value of the peak in NMR spectrum, we can obtain the value of $|M_t|$.

To calibrate the system, we adjust the phase of the NMR spectrum such that $\phi _0$ becomes the reference phase, and normalize its peak intensity as 1. Some of the experimental data of the spectra are shown in Fig.~\ref{apdx_signal} for the case of $h=0$, at $t= 0, 1.6/J$ and $6.4/J$.

By simulating the Hamiltonian evolution for different times, a range of frequency spectrum of $\left| {M_t } \right|e^{i\phi _t }$ can be obtained by the method of discrete Fourier transformation (DFT). The Fourier-transformed spectra are shown in Fig.~\ref{Fig3} for the cases of $h=0, 0.75h_c$, and $1.25h_c$, respectively. For each spectrum, totally 128 data points were collected.

\section{The precision limit of the iterative phase estimation algorithm}

In principle, it is possible to simulate the time evolution for an arbitrarily long time by mapping it back to a corresponding short time evolution. In practice, this method is limited by the precision of $J$, which is determined independently in the experiment. Here we show that when $J$ is changed by a small amount, i.e., $J \to J + \delta J$, then the error for determining the phase angle for the mapping goes as $8 n \pi \times (\delta J /J)$. In this experiment, we are able to determine the eigenvalues to the fifth digit of accuracy (see Fig.~\ref{Fig4}).

To elaborate more, let us consider the iterative phase measurement. For the moment, let use consider one of the terms $J I_z^a I_z^b$ in the Heisenberg Hamiltonian $H$ defined in Eq. (\ref{def_H}). We want to find the value of $\alpha$ such that,
\begin{equation}
e^{ - i\alpha I _z^a I _z^b}  = e^{ - i J t I_z^a I _z^b} \quad,
\end{equation}
where $t = 8n\pi /J + \tau $, and $n$ is determined by the condition that $0 < J\tau  < 8\pi$. Ideally, we have,
\begin{equation}
\alpha  = Jt = J \times \left( {8n\pi /J} \right) + J\tau  = J\tau \quad {\rm (ideal)}.
\end{equation}
If there is a fluctuation of $J \to J + \delta J$, then the wrong $\alpha$ , call it $\alpha_\delta$ , is:
\begin{equation}
\alpha_\delta= \left( {J + \delta J} \right)t = J\tau  + \delta J \times \left( {8n\pi /J + \tau } \right) \quad.
\end{equation}
The change of the phase angle, $\Delta \alpha  \equiv \alpha  - \alpha _\delta = \delta J \times t$, is therefore equal to
\begin{equation}
\Delta \alpha  = \delta J \times \left( {8n\pi /J + \tau } \right) \quad,
\end{equation}
which becomes $\Delta \alpha  \approx 8n\pi  \times \left( {\delta J/J} \right)$ for large $n$.

In the phase estimation algorithm for determining the eigenvalue $E$, if we set $E t \approx \alpha$ (up to some constant), then $\delta E \approx \Delta \alpha  /t \approx \delta J$. In this experiment, $\delta J /J \approx 0.01\%$, which makes $\delta E / E \approx 0.01 \%$. This is in agreement with the data of the ground-state and excited-state energies in Fig. \ref{Fig4}, the accuracy is about $10^{-5} \times 2 \pi J$.

On the other hand, we comment one point which may need attention in the implementation of the iterative phase estimation algorithm described in this work. In our method, although we have an accuracy about 0.04 (in units of $2 \pi J$) in reading the digit for every iteration, it is not guaranteed that the digit determined is correct for all iterations; some error-correction procedure is needed. This is because in some exceptional cases, for example, in the second iteration of the excited state, the experiment result is 0.038916 and the theoretic value is 0.039788. If, unfortunately, we obtained the experiment result as 0.041788 instead, which has a difference of about 0.002 from theoretic value, in our procedure, we would conclude that the second digit of the energy is 4, but the right answer is 3. We can only solve this problem in the following iteration; in the next iteration, even if we used the wrong value of the second digit, we will obtain a peak not lying between 0 and 1. So we can determine that the second digit should be 3 instead.

\section{Generalization to the cases of multiple eigenvalues}
In this experiment, we have chosen the case of the trial state $\left| {\psi _* } \right\rangle$ that captures two out of four eigenstates of the two-spin Hamiltonian. Therefore, we can use a single qubit (two states) to resolve the two distinct eigenvalues, and map the final state into the form defined in Eq. (\ref{fin_state}), which is then analyzed by a quantum state tomography to extract the information about the ground state $\left| {e_0 } \right\rangle$.

In general, a trial state may capture more than two eigenvalues. In this case, our procedure needs to be generalized. However, there is nothing fundamentally new, except for a more laborious repetition of the same procedures. This is the reason we decided to work on the specific case of the trial state being the linear combination of two eigenstates only.

To explain the details of how it works, we assume the ground-state energy of $H$ is unique. Define the first excited state as $\left| {e_1 } \right\rangle$. Then, any trial state can be decomposed into the following form:
\begin{equation}
\left| {\psi _* } \right\rangle  = a_0 \left| {e_0 } \right\rangle  + a_1 \left| {e_1 } \right\rangle  + a_2 \left| {e_2 } \right\rangle \quad,
\end{equation}
where $\left| {a_0 } \right|^2  + \left| {a_1 } \right|^2  + \left| {a_2 } \right|^2  = 1$, and $\left| {e_2 } \right\rangle$ represents the linear combination of all higher energy states captured by $\left| {\psi _* } \right\rangle$. Then, we perform the phase estimation algorithm, using a single probe qubit (cf. Eq. (\ref{probe_qubit})), and obtain all of the eigenvalues. Performing the same procedure for getting Eq. (\ref{fin_state}), we can obtain the following state:
\begin{equation}
\left( {b_0 \left| {e_0 } \right\rangle  + b_{20} \left| {e_2 } \right\rangle } \right)\left| 0 \right\rangle  + \left( {b_1 \left| {e_1 } \right\rangle  + b_{21} \left| {e_2 } \right\rangle } \right)\left| 1 \right\rangle ,
\end{equation}
where $\left| {b_0 } \right|^2  + \left| {b_1 } \right|^2  + \left| {b_{20} } \right|^2  + \left| {b_{21} } \right|^2  = 1$. Now, if we perform a state tomography, and extract the first part of the state, we obtain a new state
\begin{equation}
{b_0 \left| {e_0 } \right\rangle  + b_{20} \left| {e_2 } \right\rangle }
\end{equation}
which contains no eigenstate ${\left| {e_1 } \right\rangle }$. If we use this new state as the new trial state for another cycle, we get one less eigen-energy to worry about. Therefore, we can in principle eliminate the higher eigenstates one after one, and obtain the ground state in the end, using a single probe qubit.

\begin{figure}[t]
\includegraphics[width= 0.9 \columnwidth]{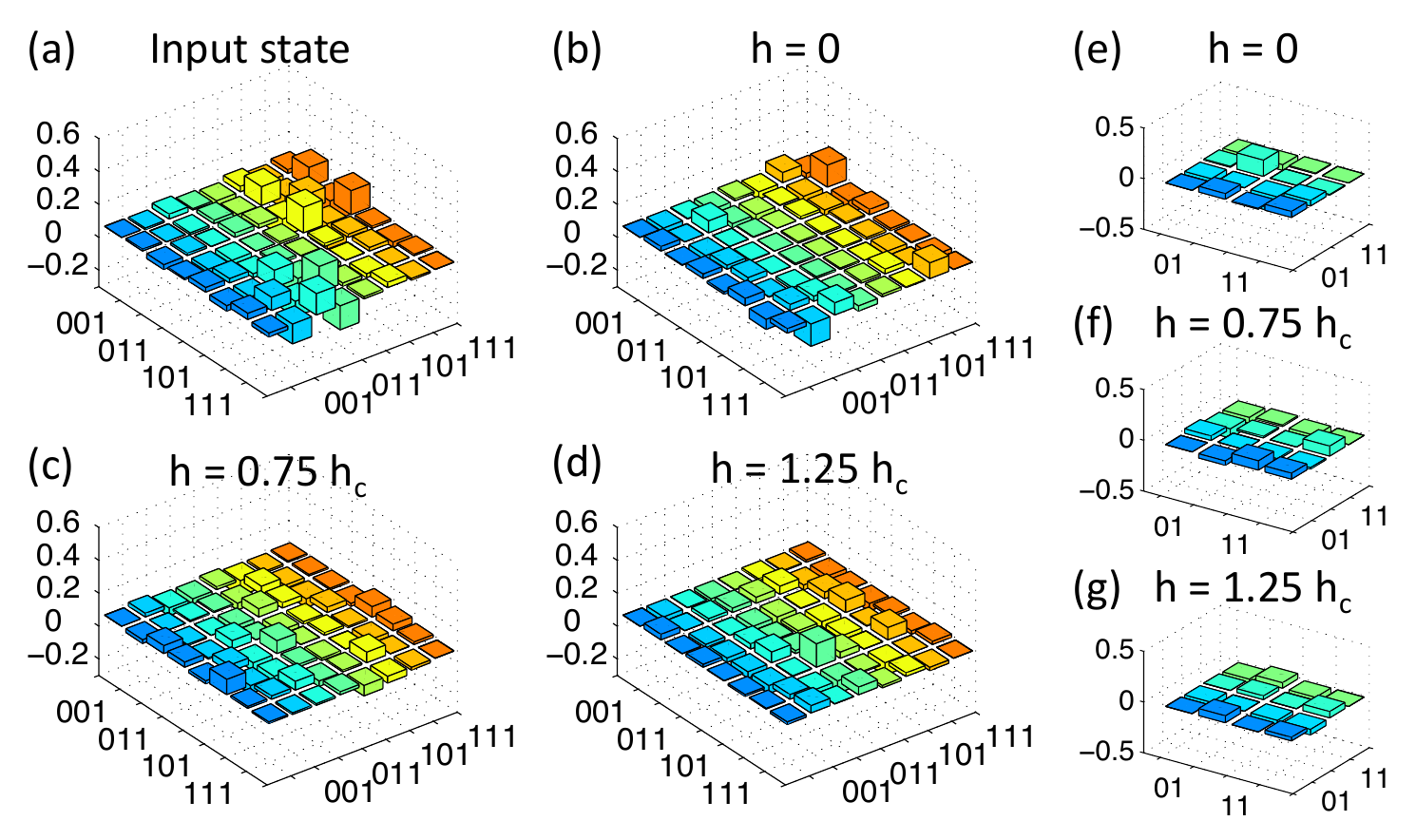}
\caption{(Color online) Imaginary parts of the tomography results. (a) the initial state $\left\vert \psi_{*} \right\rangle $. (b),(c), and (d) Three final states (Eq. (\ref{fin_state})) for the cases, respectively, $h=0$, $h=0.75 h_c$, and $h=1.25 h_c$. (e), (f), and (g) The first $4\times4$ part of each density matrix above.}\label{apdx_imag}
\end{figure}

\begin{figure}[t]
\includegraphics[width=0.9 \columnwidth]{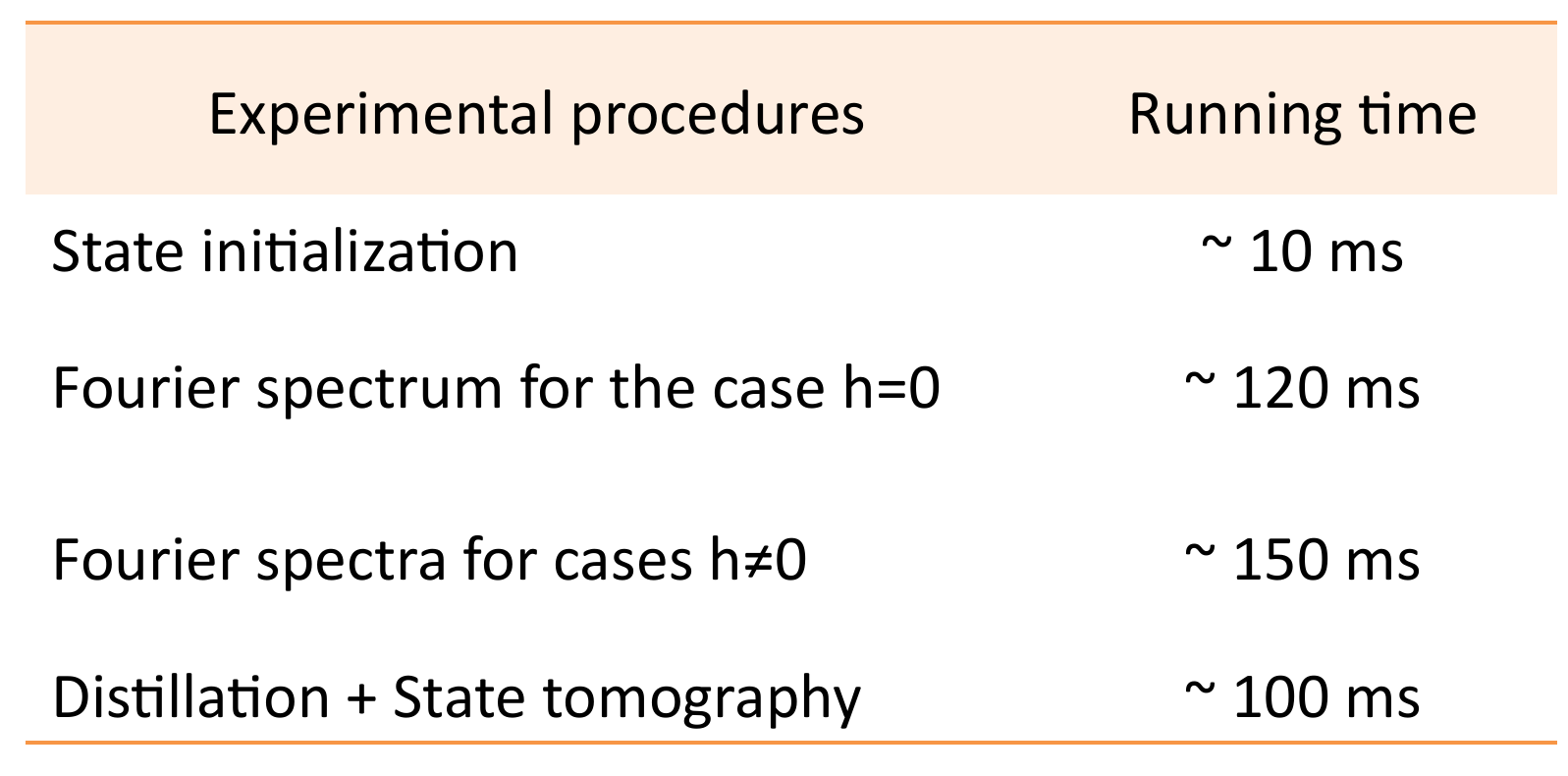}
\caption{(Color online) Running times for various experimental procedures. The time scale is taken from the longest ones. Distillation refers to the procedure to obtain Eq. (\ref{fin_state}).}\label{apdx_times}
\end{figure}

\section{Supplementary data}

\begin{enumerate}
\item In Fig.~\ref{apdx_imag}, the imaginary parts of the results from the quantum state tomography are shown.
\item In Fig.~\ref{apdx_times}, the running times of various experimental procedures are shown.
\end{enumerate}

%


\end{document}